\documentclass[aps,10pt,twocolumn,prb,groupedaddress,showpacs]{revtex4}

\usepackage{graphicx}
\usepackage[caption=false]{subfig}
\usepackage{amsmath}
\usepackage{bbold}
\usepackage{latexsym,color}

\begin{document}


\title{Majorana bound states in magnetic skyrmions }


\author{Guang Yang,$^1$ Peter Stano,$^1$ Jelena Klinovaja,$^2$ and Daniel Loss$^{1,2}$}
\affiliation{$^1$RIKEN Center for Emergent Matter Science, Wako, Saitama 351-0198, Japan \\
$^2$Department of Physics, University of Basel, Klingelbergstrasse 82, CH-4056 Basel, Switzerland}



\date{\today}

\begin{abstract}
Magnetic skyrmions are highly mobile nanoscale topological spin textures. We show, both analytically and numerically, that a magnetic skyrmion of an even azimuthal winding number placed in proximity to an $s$-wave superconductor hosts a zero-energy Majorana bound state in its core, when the exchange coupling between the itinerant electrons and the skyrmion is strong. This Majorana bound state is stabilized by the presence of a spin-orbit interaction. We propose the use of a superconducting tri-junction to realize non-Abelian statistics of such Majorana bound states.
\end{abstract}



\maketitle


\section{Introduction}
 
Recently, there has been considerable interest in magnetic skyrmions, \cite{skyrev} particle-like topological spin textures discovered in chiral ferromagnets with Dzyaloshinskii-Moriya interaction and in dipolar ferromagnets with uniaxial anisotropy. These magnetic nanostructures are objects of many internal degrees of freedom and can be driven by ultralow electric current densities. \cite{drive1,drive2} Experimentally, skyrmions with both odd and even azimuthal winding numbers have been observed. \cite{sky1,sky2,bisky} The latter exist in dipolar ferromagnets where a ``spin helicity'' degree of freedom allows for the formation of more complex structures. In thin film samples, magnetic skyrmions can be stabilized over a wide temperature range \cite{yu11,garel82} including near the absolute zero. 

In this work, we demonstrate the potential application of magnetic skyrmions to topological quantum computation (TQC). \cite{kitaev03,nayak08} We investigate a magnetic skyrmion in proximity to a conventional $s$-wave superconductor
and show that for strong exchange coupling between the itinerant electrons and the skyrmion there exists a zero-energy Majorana bound state (MBS) in the skyrmion core, if the skyrmion has an even azimuthal winding number. These MBSs exhibit non-Abelian statistics, and can be braided using superconducting tri-junctions.

There are two issues concerning TQC with MBSs in magnetic skyrmions. For a skyrmion with a single spin-flip in radial direction, the MBS localization length is comparable to the skyrmion size, leading to hybridization with gapless bulk states. This can be prevented by using skyrmions with multiple spin-flips radially, or by introducing a spin-orbit interaction (SOI), both of which stabilize the MBS. 
Second, the MBS is generically accompanied by subgap localized fermionic states. Fortunately, these subgap fermions are spatially separated from the MBS due to a ``centrifugal'' force, and in addition, they respond to electric and magnetic fields, allowing further discrimination from the MBS.

In one dimension, it is known that a helical field or spin order, combined with the proximity-induced superconductivity, leads to the emergence of MBSs. \cite{gangadharaiah11L,klinovaja_stano12,kjaergaard12,klinovaja13L,braunecker13L,vazifeh13L,nadj13,pientka13,nadj14,kim2015:PRB,pawlak15} 
This lies in the fact that a helical field is gauge equivalent \cite{braunecker10} to a uniform Zeeman field and a Rashba SOI, the latter two being the essential ingredients of topological phases supporting MBSs. 
Qualitatively, our results can be understood through an analogy between a magnetic skyrmion and a 1D helical spin order in radial direction.
A skyrmion lattice in two dimensions was found to induce an effective $p$-wave pairing, resulting in degenerate zero-energy bound states. \cite{nakosai13} Similar pairing was deduced in a toy model of a single skyrmion in Ref.~\onlinecite{chen}, while  Shiba states bound to a  skyrmion were found in Ref.~\onlinecite{pershoguba}. 

This paper is organized as follows. In Sec.~II, we introduce the model which captures the physics at the interface between a magnetic skyrmion and an $s$-wave superconductor. We show that under certain conditions a MBS emerges near the core of the skyrmion of an even azimuthal winding number and obtain the MBS wave function. In Sec.~III, we study the quasiparticle spectrum of the system and discuss the robustness of the MBSs. In Sec.~IV, we perform a tight-binding calculation whose results further support the findings in Secs.~II and III. In Sec.~V, we discuss the realization of non-Abelian statistics of the MBSs. We conclude in Sec.~VI.

\section{Model}
 We consider a single magnetic skyrmion with a core of uniformly polarized spins. Outside the core region, the skyrmion spin varies locally in both radial and azimuthal directions until the far asymptotic region, again with uniform spin polarization. We parametrize the skyrmion spin texture as 
\begin{equation}
\label{y1}
\hat{N}(\mathbf{r})=\big(\sin f(r)\cos n\theta, \sin f(r)\sin n\theta, \cos f(r)\big)
\end{equation}
in polar coordinates $\mathbf{r}=(r\cos\theta,r\sin\theta)$, where the piecewise function $f(r)$ is defined as: 0 for $r<r_0$, $\pi(r-r_0)/R$ for $r_0\leq r\leq r_0+pR$ and $\pi$ for $r>r_0+pR$, with $p$ a positive integer and $r_0\ll R$. By construction, the skyrmion core has size $2r_0$ and the skyrmion spin flips $p$ times moving radially outwards from the center. The azimuthal winding number $n$ takes integer values.

The Hamiltonian for itinerant electrons exchange coupled to the skyrmion spin texture is
\begin{equation}
\label{y2}
H_0=\int d\mathbf{r}\, \sum_{\gamma,\delta}\psi^{\dagger}_\gamma(-\frac{\nabla^2}{2m}-\mu+\alpha \hat{N}\cdot \vec{\sigma})_{\gamma\delta}\psi_\delta,
\end{equation}
where $\psi_\gamma$ is the annihilation operator of electron with spin $\gamma=\uparrow, \downarrow$, $m$ is the effective mass, $\mu$ is the chemical potential, $\alpha$ is the exchange coupling constant, and $\vec{\sigma}$ are Pauli matrices acting in spin space. (We set $\hbar=1$ throughout.)
The proximity-induced superconductivity is described by $H_S= \int d\mathbf{r} (\Delta \psi_{\uparrow}^{\dagger} \psi_{\downarrow}^{\dagger} +\textrm{H.c.})$, where $\Delta=\Delta_0 e^{i \varphi}$ is the pairing potential. The full Hamiltonian is $H=H_0+H_S=\int d \mathbf{r}\Psi^{\dagger}\mathcal{H}\Psi/2$, where
\begin{equation}
\label{y3}
\mathcal{H}=(-\frac{\nabla^2}{2m}-\mu)\tau_z+\alpha \hat{N}\cdot \vec{\sigma}+\Delta\tau_++\Delta^*\tau_-
\end{equation}
in Nambu basis $\Psi^{\dagger}=[\psi_{\uparrow}^{\dagger},\psi_{\downarrow}^{\dagger},\psi_{\downarrow},-\psi_{\uparrow}]$. The Pauli matrices $\vec{\tau}$ act in particle-hole space, and $\tau_{\pm}=(\tau_x\pm i \tau_y)/2$. 

Quasiparticle excitations above the superconducting ground state are described by operators $\chi^{\dagger}=\int d \mathbf{r} \sum_{\gamma}  ( u_{\gamma}\psi_{\gamma}^{\dagger}+v_{\gamma}\psi_{\gamma})$ satisfying the Bogoliubov-de Gennes (BdG) equation $\mathcal{H} \Upsilon(\mathbf{r})=E\Upsilon(\mathbf{r})$, where $\Upsilon(\mathbf{r})=[u_{\uparrow}(\mathbf{r}),u_{\downarrow}(\mathbf{r}),v_{\downarrow}(\mathbf{r}),-v_{\uparrow}(\mathbf{r})]^T$. 
We look for zero-energy solutions to the BdG equation satisfying the Majorana condition $\mathcal{C}\Upsilon(\mathbf{r}) = \eta \Upsilon(\mathbf{r})$, where the particle-hole operator $\mathcal{C}=\sigma_y \tau_y K$, with $K$ the complex conjugation, and $\eta$ is some constant. The BdG equation is solved by eigenstates of the angular momentum-like operator $-i \partial_\theta + (n/2) \sigma_z$ commuting with $\mathcal{H}$,
\begin{equation}
\label{y4}
\Upsilon^l(r,\theta)=e^{i(l-\frac{n}{2}\sigma_z)\theta}e^{i\frac{1}{2}\tau_z\varphi} \Upsilon^l(r)
\end{equation}
with eigenvalues $l$, where the radial wave functions $\Upsilon^l(r)=[u_{\uparrow}^l(r),u_{\downarrow}^l(r),v_{\downarrow}^l(r),-v_{\uparrow}^l(r)]^T$. A single-valued wave function requires $l$ be an integer (half-integer) for even (odd) $n$. Under particle-hole transformation, solutions with angular momentum $l$ transform into those with $-l$. The quasiparticle spectrum is thus symmetric with respect to the $l=0$ sector, where non-degenerate zero-energy solutions must reside. The zero-mode wave function is $\Upsilon^0(r,\theta)=e^{-i\frac{n}{2}\sigma_z\theta}e^{i\frac{1}{2}\tau_z\varphi} \Upsilon^0(r)$, where $\Upsilon^0(r)$ is the kernel of the real matrix operator 
\begin{equation}
\label{y5}
\mathcal{H}^l(r)=(-\frac{[\nabla^2]_r}{2m} -\mu)\tau_z +\alpha \sigma_z \cos f +\alpha  \sigma_x \sin f+\Delta_0 \tau_x
 \end{equation}
at $l=0$, with $[\nabla^2]_r=\partial_r^2+\frac{1}{r}\partial_r-\frac{1}{r^2}(l-\frac{n}{2}\sigma_z)^2$.
Without loss of generality, we choose $\Upsilon^0(r)$ to be real. The Majorana condition then translates to $v_{\uparrow, \downarrow}^0=\eta u_{\uparrow, \downarrow}^0$ with $\eta=\pm1$, which allows for the reduction of the radial equation $\mathcal{H}^{l=0}(r)\Upsilon^0(r)=0$ to
\begin{align}
\label{y6}
\left( \begin{array}{cc}
-\frac{[\nabla^2]_r}{2m} -\mu+\alpha \cos f & \alpha \sin f +\eta\Delta_0\\
\alpha \sin f -\eta\Delta_0 & -\frac{[\nabla^2]_r}{2m} -\mu-\alpha \cos f
 \end{array} \right) \Phi=0
 \end{align}
 in terms of the two-spinor $\Phi(r)\equiv[u_{\uparrow}^0(r),u_{\downarrow}^0(r)]^T$. In the following, we solve Eq.~(\ref{y6}) for $n=2$,  which is the simplest topologically stable spin configuration for an even $n$.
 
 It is useful to make a rotation of Eq.~(\ref{y6}) by the unitary operator $U(r)=e^{i\frac{1}{2}\sigma_yf(r)}$, which gives
 \begin{widetext}
 \begin{align}
 \label{y7}
\left( \begin{array}{cc}
-\frac{1}{2m} (\partial_r^2+\frac{1}{r}\partial_r-\frac{1}{r^2})-\tilde{\mu}+\alpha  & i\frac{f'}{2m} \hat{p}_r+\frac{f''}{4m}+\eta\Delta_0 \\
-i\frac{f'}{2m} \hat{p}_r-\frac{f''}{4m} -\eta\Delta_0 & -\frac{1}{2m} (\partial_r^2+\frac{1}{r}\partial_r-\frac{1}{r^2})-\tilde{\mu}-\alpha 
 \end{array} \right) \tilde{\Phi}=0,
  \end{align}
 \end{widetext}
where $\tilde{\Phi}(r)\equiv U(r)\Phi(r)=[\tilde{u}_{\uparrow}(r), \tilde{u}_{\downarrow}(r)]^T$, $\tilde{\mu}=\mu-f'^2/8m$ and the Hermitian radial momentum operator $\hat{p}_r=-i(\partial_r+1/2r)$. \cite{fujikawa08} It is easy to verify that the full wave function after the rotation satisfies the Majorana condition.
Equation~(\ref{y7}) shows that the spatially varying skyrmion spin texture, where $f'\neq 0$, renormalizes the chemical potential, and more importantly, generates an effective SOI, \cite{footnote2} thereby establishing the connection to a 1D system with helical spin order. \cite{gangadharaiah11L,klinovaja_stano12,kjaergaard12,klinovaja13L,braunecker13L,vazifeh13L,nadj13,pientka13,nadj14,kim2015:PRB,pawlak15,braunecker10}

\begin{figure}
\centering
\includegraphics[width=3.2in]{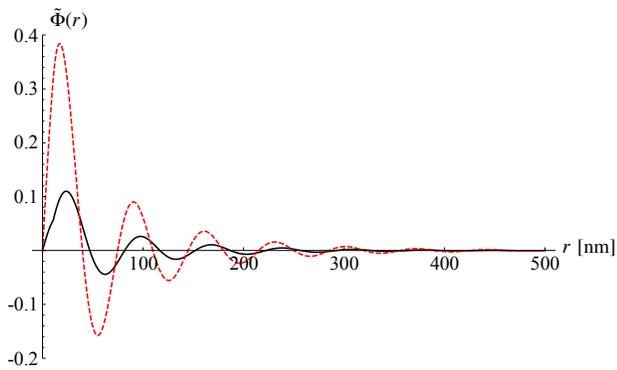}
\caption{(Color online) Radial wave function $\tilde{\Phi}(r)=[\tilde{u}_{\uparrow}(r), \tilde{u}_{\downarrow}(r)]^T$ of the MBS $\chi^{\textrm{in}}$, with the solid (black) and dashed (red) lines denoting $\tilde{u}_{\uparrow}(r)$ and $\tilde{u}_{\downarrow}(r)$, respectively. We set $\mu=0$, $\alpha=1$~meV, $\Delta_0=0.5$~meV, $R=100$~nm, $r_0=10$~nm, $n=2$, and $pR$ much larger than the shown range. The discontinuities in $\tilde{\Phi}'(r)$ at $r=r_0$ \big(less discernible for $\tilde{u}_{\downarrow}(r)$\big) arise from singularities in $f''(r)$, Eq.~(\ref{y7}).}
\end{figure}

%
%

We solve Eq.~(\ref{y7}) by exploring the similarity between our system and a 1D topological superconductor (TSC). We look for two Majoranas $\chi^{\textrm{in}}$ and $\chi^{\textrm{out}}$, one located near the inner boundary $r=r_0$ and the other located near the outer boundary $r=r_0+pR$. These are the analogues of the Majorana end states in a 1D TSC. In Appendix A, we show detailed construction of the analytical wave functions of $\chi^{\textrm{in}}$ and $\chi^{\textrm{out}}$, as well as a comparison with results from exact numerical diagonalization of the Hamiltonian in Eq.~(\ref{y5}). The analytical and numerical solutions agree well. We find that for $\alpha^2>\tilde{\mu}^2+\Delta_0^2$ a zero-energy MBS $\chi^{\textrm{in}}$ exists, an example plotted in Fig.~1. This MBS is accompanied with a \emph{delocalized} (extending to infinity) Majorana mode $\chi^{\textrm{out}}$. Fixing other parameters and decreasing the exchange energy, $\chi^{\textrm{in}}$ delocalizes and turns into an extended state at the critical point when $\alpha^2=\tilde{\mu}^2+\Delta_0^2$ (we numerically confirm this analytical result for the transition, corresponding also to the closing and reopening of a spectral gap). The inner and outer Majoranas then hybridize into a finite-energy fermion. Away from the critical point, $\chi^{\textrm{in}}$ and $\chi^{\textrm{out}}$ can still have a non-zero overlap in a finite-size skyrmion. The ratio $\xi/pR$, with $\xi$ the decay length of $\chi^{\textrm{in}}$, characterizes the protection of the MBS from hybridization. A rough upper bound for $\xi$ is given by $v_F/\Delta_{\min}$, where $v_F$ is the Fermi velocity set by the larger of the exchange energy $\alpha$ and $E_{\lambda}=m\lambda^2/2$, where $\lambda=\pi/2mR$ is the strength of the effective SOI generated by the skyrmion, and $\Delta_{\min}=\min\{\Delta_0,|\alpha-\sqrt{\Delta_0^2+\tilde{\mu}^2}|,\Delta_p\}$, where $\Delta_p=2\Delta_0\sqrt{E_{\lambda}(1+\tilde{\mu}/\alpha)/\alpha}$ is the effective $p$-wave pairing gap. \cite{alicea10} We find $\xi\sim R$ in the realistic parameter regime $\alpha>\Delta_0>E_{\lambda}$. Thus, a skyrmion with $p=1$ cannot effectively localize a MBS even in the topological phase when $\alpha^2>\tilde{\mu}^2+\Delta_0^2$. In practice, a pulse laser or a large field gradient may be used to excite skyrmions with $p>1$, \cite{yucomm} hosting well-localized MBSs. Given an even number of such skyrmions, the extended outer Majoranas hybridize with each other and drop out of the ground state degeneracy, leaving only the MBSs which can be used for TQC.

\section{Quasiparticle spectrum}
We obtain the quasiparticle spectrum by exact numerical diagonalization of the Hamiltonian in Eq.~(\ref{y5}). For simplicity, we let $r_0= 0$ and consider a skyrmion in a finite region of size $L$, requiring that the wave functions vanish for $r \geq L$. For large $p$, we expect to find two MBSs  in the topological phase, one localized near $r=0$, as shown in Fig.~2(d), and the other localized near $r=pR$. The localization length $\xi$ of the MBSs depends on the skyrmion-generated SOI. We have confirmed numerically (not shown), that $\xi \propto \lambda$ in the strong SOI regime where $E_{\lambda}\gg \alpha$ and $\xi \propto 1/\lambda$ in the weak SOI regime with $E_{\lambda} \ll \alpha$. Exactly the same behavior was found for 
MBSs in 1D TSCs. \cite{JL12}

\begin{figure}
\centering
\includegraphics[width=0.5\textwidth]{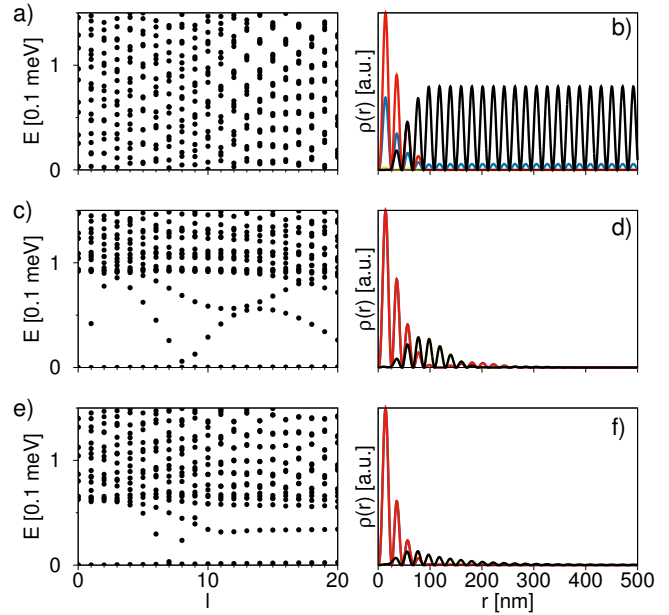}
\caption{(Color online) Excitation spectrum (left panels) and the probability density $\rho(r)=r|\Upsilon^0(r)|^2$ of the lowest eigenstate in the $l=0$ sector (right panels, with the $u_{\uparrow}^0,u_{\downarrow}^0,v_{\uparrow}^0$, and $v_{\downarrow}^0$ components shown in yellow, blue, black, and red, respectively), for a system size $L=1\,\mu$m with Dirichlet boundary conditions and for realistic parameters $m=m_e$, $\mu=0$, $\alpha=1$~meV, $\Delta_0=0.5$~meV, $\varphi=0$, $R=100$~nm, $r_0=0$, and $n=2$.
(a-b) Skyrmion with $p=1$. (c-d) Skyrmion with $p=10$. (e-f) Skyrmion with $p=1$ in the presence of a SOI described by $H_{so}$, with $l_{so}=100$~nm. The energy of the lowest eigenstate in the $l=0$ sector returned by numerics is 3240~neV in (a), 20~neV in (c), and 201~neV in (e). Excitation energies for $l<0$ are not plotted, being identical to those for $l>0$. In right panels, wave functions are plotted for $r\leq L/2$, to be separated from the degenerate state localized at the outer boundary. In (d) and (f), $|v_{\uparrow, \downarrow}^0|=| u_{\uparrow, \downarrow}^0|$ and only the hole components are plotted. }
\end{figure}

Figure~2 shows the spectrum in the topological phase. For a skyrmion with a single radial spin-flip, Fig.~2(a), there is no spectral gap and the MBSs hybridize, see Fig.~2(b). For a skyrmion with multiple radial spin-flips, a gap separating the MBSs at $l=0$ and the quasi-continuum levels can be identified. Inside the gap, we find two sets of localized fermionic states with finite angular momenta $l$, \cite{footnote} associated with the two MBSs. The localized states near the outer MBS have nearly zero energies and form an almost flat (yet distinctively quadratic) band, while those near the inner MBS have higher energies and form a more dispersive band. For the purpose of TQC, we are concerned with the subgap states near the inner MBS. Although these localized fermions cannot change the nonlocal fermion parity shared by two spatially separated MBSs (in two different skyrmions), \cite{akhmerov10} they cause dephasing and affect the signal strength at the readout. The level spacing of these states thus sets a bound for the allowed temperature fluctuation when measurement is being carried out. Interestingly, we find that these states have nonzero charge and spin expectation values (see Appendix B), which may allow distinguishing them experimentally from the inner MBS, e.g., by applying electric and/or magnetic fields (and also via transport experiments \cite{ktlaw}). In contrast, the subgap states near the outer MBS have almost zero charge and spin expectations. Our numerics also shows that all subgap states are subject to a ``centrifugal'' force due to their finite angular momenta.  At increasing $l$, they move away from the skyrmion core, and thus from the inner MBS. In practice, quasiparticle poisoning may also arise from the electrical driving of skyrmions. However, similar issues occur for the manipulation of MBSs in quantum wires as well, \cite{rainis12,klepj, pedrocchi15} and in principle a continuous error correction is necessary for performing TQC. \cite{hutter15} Refs.~\onlinecite{goldstein12,yang14,kells15} discussed general approaches to optimization of TQC in the presence of quasiparticle poisoning.


We now consider a magnetic skyrmion with a single radial spin-flip in the presence of an extrinsic SOI.
For numerical convenience (to be able to reduce the problem to a 1D one), we consider SOI of a special form, $H_{ so}=\int d\mathbf{r}\,\psi^{\dagger}\left[\{\cos\theta, ({\vec\sigma}\times \vec{p})\cdot \hat{z} \}+\{\sin\theta,(\vec{\sigma}\cdot \vec{p})\} \right]\psi/4ml_{so}$, \cite{sato09} where $\{,\}$ is the anticommutator and $l_{so}$ is chosen such that the extrinsic and skyrmion-generated SOIs are comparable in strength. As seen in Figs.~2(e) and (f), the extrinsic SOI opens a spectral gap and stabilizes the MBSs, though also with the appearance of subgap localized fermions.  

\section{Tight-binding model}
 The existence of MBSs in magnetic skyrmions is further supported by a numerical tight-binding calculation. 
Fig.~3 shows the spectrum and the probability density profile of the zero-energy state for a skyrmion with $n=2$ and $p=25$.
The skyrmion induces many fermionic subgap states, but still
the zero-energy MBSs are clearly seen, well separated from the other states, similar to Figs.~2(c) and (e). These results support our previous conclusions based on analytics and 1D numerics.


\begin{figure}
\centering
\includegraphics[width=3.2in]{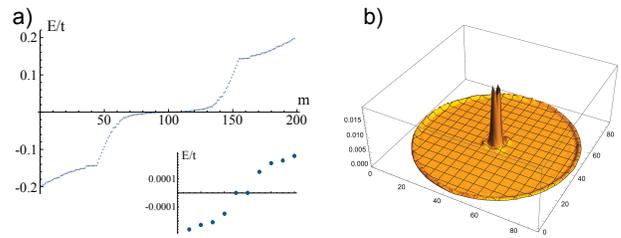}
\caption{(Color online) (a) Energy spectrum of $\mathcal{H}$ found numerically in the tight-binding model with hopping amplitude $t$ and lattice constant $a$, in the presence of a skyrmion with $n=2$ and $p=25$. Here, $m$ labels the eigenstates. The inset shows ten energies closest to the chemical potential.
 (b) The probability density profile of the lowest  eigenstate at near-zero energy $E_M/t=1.6 \times 10^{-7}$. This  electron state and its hole partner at $-E_M$ can be identified with two weakly overlapping MBSs (one at the center and one at the system edge) and are very well separated in energy from the remaining states. We have used $\alpha/t=1.2 $, $\Delta_0/t=0.4$, $\mu/t=0$, and $R/a=3.5$.}
\end{figure}

\section{Non-Abelian statistics}

 Non-Abelian statistics does not follow immediately from braiding the MBSs in magnetic skyrmions, which unlike the MBSs in $p+ip$ superconductors \cite{volovik99,read00,ivanov01} are not bound to superconducting vortices. We overcome this difficulty with the help of a superconducting tri-junction, \cite{fu2008} in a spirit similar to TQC in one dimension. \cite{tqc1d} The tri-junction divides the space into three parts, with order-parameter phases $\varphi=\varphi_1,\varphi_2,\varphi_3$. As shown earlier, the $\varphi$-phase enters the MBS wave function as $\Upsilon^0(\mathbf{r})\propto e^{\frac{1}{2}\tau_z\varphi}$. A branch cut is needed to avoid the multi-valuedness of the MBS wave function. Consider now two MBSs $\chi^A$ and $\chi^B$, initially located in the $\varphi_1$-region and $\varphi_2$-region, respectively. The exchange of $\chi^A$ and $\chi^B$ involves three steps, as sketched in Fig.~4(a): First, $\chi^A$ crosses the $\varphi_1$-$\varphi_3$ junction; second, $\chi^B$ crosses the $\varphi_1$-$\varphi_2$ junction; and third, $\chi^A$ crosses the $\varphi_2$-$\varphi_3$ junction. When a MBS crosses a junction, the $\varphi$-phase locally felt by the MBS changes, giving rise to a rotation of the phase of the wave function. Such a phase rotation may be clockwise or anticlockwise, as represented in the $\varphi$-plane. However, phase rotations with the formation of intermediate $\pi$-junctions do not correspond to physical manipulations of MBSs, since $\pi$-junctions introduce additional zero modes \cite{fu2008,sau10L1} that were not present in the original problem. Let $ \varphi_{ij}=\min\{|\varphi_i-\varphi_j|,2\pi-|\varphi_i-\varphi_j|\}$, where $i,j=1,2,3$. The case $\{\varphi_{12}+\varphi_{23},\varphi_{23}+\varphi_{31},\varphi_{31}+\varphi_{12} \}>\pi$ corresponds to the situation where the three phases $\varphi_1,\varphi_2,\varphi_3$ form a $Y$-shape in the $\varphi$-plane, as shown in Fig.~4(b), such that all physical phase rotations are in the same direction. This leads to the non-Abelian transformation rule of the MBSs after the exchange: $\chi^A\rightarrow -\nu\chi^B$ and $\chi^B\rightarrow \nu\chi^A$, with $\nu=\pm1$. In practice, arrays of tri-junctions may be constructed to implement TQC.  


\begin{figure}
\centering
\includegraphics[width=3.2in]{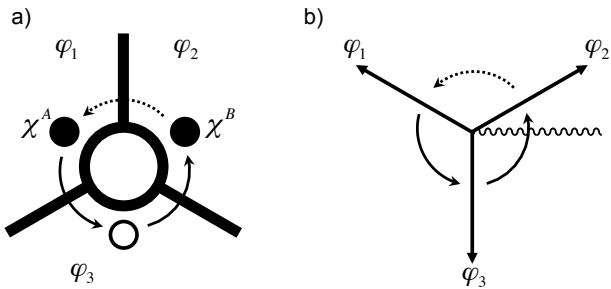}
\caption{Braiding of two MBSs $\chi^A$ and $\chi^B$ (solid circles): (a)~real-space manipulation, with the hollow circle indicating the temporary position of $\chi^A$, and (b)~corresponding phase rotations in the MBS wave function, with the wavy line denoting a branch cut. The solid and dashed arrows show the trajectories of $\chi^A$ and $\chi^B$, respectively.}
\end{figure}

 We note that the above scheme of TQC relies on the existence of a bulk gap in the system, i.e., in the far asymptotic region between skyrmions. Otherwise, braiding of MBSs can lead to non-universal results. Although not explicitly included in the model, we assume that such a bulk gap can be provided by the SOI ubiquitous in the magnetic materials hosting skyrmions.

\section{Conclusions}
 We have demonstrated the existence of MBSs in magnetic skyrmions with even azimuthal winding numbers, placed in proximity to an $s$-wave superconductor. The electrical drivability of magnetic skyrmions makes the real-space manipulation of MBSs straightforward. TQC can be performed with the help of superconducting tri-junctions.

\begin{acknowledgments}
We thank X. Z. Yu and Y. Tokura for helpful discussions. We acknowledge partial support from the Swiss NSF and NCCR QSIT. This work was supported by JSPS KAKENHI Grant Number 16H02204. 
\end{acknowledgments}


\appendix

\section{Construction of analytical Majorana wave functions}

In this appendix, we construct analytical wave functions for the Majoranas $\chi^{\textrm{in}}$ and $\chi^{\textrm{out}}$ by solving Eq.~(\ref{y7}). 
For $r<r_0$, the equation is solved by 
\begin{equation}
[\tilde{u}_{\uparrow}(r), \tilde{u}_{\downarrow}(r)]^T=J_1(\kappa r)[a_{\uparrow},a_{\downarrow}]^T,
\label{WFcore}
\end{equation}
where $J_1(x)$ is the Bessel function of the first kind and the constants $a_{\uparrow},a_{\downarrow}$ satisfy 
\begin{align}
\label{y8}
\left( \begin{array}{cc}
\frac{\kappa^2}{2m}  -\mu+\alpha  & \eta\Delta_0\\
-\eta\Delta_0 & \frac{\kappa^2}{2m}  -\mu-\alpha 
 \end{array} \right) 
 \left( \begin{array}{c}
 a_{\uparrow} \\
 a_{\downarrow}
\end{array} \right)=0.
 \end{align}
For non-trivial solutions, we require
\begin{equation}
\label{y9}
(\frac{\kappa^2}{2m} -\mu)^2-\alpha^2+\Delta_0^2=0.
 \end{equation}
Let us focus on the regime of strong exchange coupling, where $\alpha>\Delta_0$. The roots of Eq.~(\ref{y9}) are then classified as: $\kappa=\pm \kappa_1, \pm i\kappa_2$, for $\alpha^2>\mu^2+\Delta_0^2$; $\kappa=\pm \kappa_1, \pm \kappa_2$, for $\alpha^2<\mu^2+\Delta_0^2$ and $\mu>0$; and $\kappa=\pm i\kappa_1, \pm i\kappa_2$, for $\alpha^2<\mu^2+\Delta_0^2$ and $\mu<0$. Here, $\kappa_1,\kappa_2$ are positive real numbers. In all cases, we find two solutions that are regular at the origin, for both $\eta=\pm1$. These solutions are used to construct the wave function of $\chi^{\textrm{in}}$. 
For $r_0\leq r\leq r_0+pR$, we find convergent power series solutions \cite{sau10L1}
\begin{equation}
[\tilde{u}_{\uparrow}(r), \tilde{u}_{\downarrow}(r)]^T=(e^{-\kappa r}/\sqrt{r})\sum_{s=0}^{\infty}(r_0/r)^s[b_{\uparrow}^s ,b_{\downarrow}^s ]^T.
\label{WF}
\end{equation} 
The lowest-order coefficients $b_{\uparrow}^0, b_{\downarrow}^0$ satisfy
\begin{align}
\label{y10}
\left( \begin{array}{cc}
-\frac{\kappa^2}{2m} -\tilde{\mu}+\alpha  &-\lambda\kappa+ \eta\Delta_0 \\
\lambda\kappa -\eta\Delta_0 & -\frac{\kappa^2}{2m} -\tilde{\mu}-\alpha 
 \end{array} \right) 
 \left( \begin{array}{c}
 b_{\uparrow}^0 \\
 b_{\downarrow}^0
\end{array} \right)=0,
 \end{align}
where $\lambda=\pi/2mR$. Eq.~(\ref{y10}) is solved by requiring
\begin{equation}
 \label{y11}
(\frac{\kappa^2}{2m} +\tilde{\mu})^2-\alpha^2+(\lambda\kappa-\eta\Delta_0)^2=0.
 \end{equation}
We have verified the existence of recursion relations relating $b_{\uparrow}^0, b_{\downarrow}^0$ to higher-order coefficients $b_{\uparrow}^s, b_{\downarrow}^s$ so the series solutions are indeed well-defined. The definition of $\chi^{\textrm{in}}$ instructs us to select  solutions that decay exponentially with distance outside the inner boundary $r=r_0$, with $\textrm{Re}\kappa>0$. Similarly, for $\chi^{\textrm{out}}$ we look for solutions that grow exponentially with distance inside the outer boundary $r=r_0+pR$, with $\textrm{Re}\kappa<0$. By carefully analyzing the structure of Eq.~(\ref{y11}), we find three exponentially decaying (growing) solutions for $\eta=-1$ ($\eta=1$), when $\alpha^2>\tilde{\mu}^2+\Delta_0^2$, and two such solutions for both $\eta=\pm1$, when $\alpha^2<\tilde{\mu}^2+\Delta_0^2$. These solutions can all be made real under proper linear combinations and are used to construct the wave functions of $\chi^{\textrm{in}}$ and $\chi^{\textrm{out}}$. In the far asymptotic region $r>r_0+pR$, the solutions to Eq.~(\ref{y7}) resemble those inside the skyrmion core. To obtain a bound state for $\chi^{\textrm{out}}$, we need solutions that are normalizable as $r\rightarrow \infty$. By above analysis, there are two such solutions for $\alpha^2<\mu^2+\Delta_0^2$ and $\mu<0$, and at most one in other cases. 

The boundary conditions at $r=r_0$ impose four constraints on the radial wave function $\tilde{\Phi}(r)$ of $\chi^{\textrm{in}}$ and its derivative. Combined with the normalization condition, the total of five constraints equals the number of independent solutions available for constructing $\tilde{\Phi}(r)$, when $\alpha^2>\tilde{\mu}^2+\Delta_0^2$. (Note that solutions corresponding to different $\eta$ values should not be mixed, as they transform in different ways under particle-hole operator $\mathcal{C}$.) In this case, a non-degenerate zero-energy MBS is uniquely determined, as shown in Fig.~1. When $\alpha^2<\tilde{\mu}^2+\Delta_0^2$, the number of constraints exceeds that of the solutions with desired asymptotics and a MBS does not exist. 

The wave function of $\chi^{\textrm{out}}$ is obtained in a similar way. Again, five independent solutions to Eq.~(\ref{y7}) are needed to construct a bound state through matching the boundary conditions. This assumes the parameter domain $\tilde{\mu}^2+\Delta_0^2<\alpha^2<\mu^2+\Delta_0^2$ and $\mu<0$, which is unrealistic since by definition $\tilde{\mu}<\mu$. We conclude that a MBS can never be established for $\chi^{\textrm{out}}$, although a non-bound-state Majorana is indeed possible. The delocalization of $\chi^{\textrm{out}}$ is understood by noticing that the far asymptotic region becomes gapless when $\alpha^2>\mu^2+\Delta_0^2$, which for negative $\mu$ precedes the occurrence of the topological phase. For $\mu>0$, the far asymptotic region has a small gap when $\tilde{\mu}^2+\Delta_0^2<\alpha^2<\mu^2+\Delta_0^2$, which however is incapable of localizing a bound state for $\chi^{\textrm{out}}$. 

We have also obtained the Majorana wave functions using exact numerical diagonalization of the Hamiltonian in Eq.~(\ref{y5}). In Fig.~5, we compare the probability density of the MBS $\chi^{\textrm{in}}$, obtained from analytical approach (using Eqs.~\eqref{WFcore} and \eqref{WF} in the lowest order) and from numerics. The two curves agree well.

\begin{figure}
\centering
\includegraphics[width=3in]{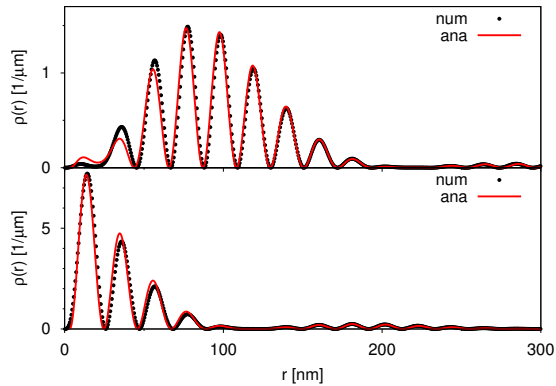}
\caption{Comparison of analytical and numerical solutions for $u_{\uparrow}^0$ (upper panel) and $u_{\downarrow}^0$ (lower panel) components of the probability density $\rho(r)=r |\Phi(r)|^2$ of the MBS $\chi^{\textrm{in}}$ in Fig.~2(d).}
\end{figure}

Fixing other parameters and varying the exchange energy, the system can be driven from the topological phase to the non-topological phase. During this process, the spectral gap closes and reopens, while the MBS
delocalizes and eventually disappears from the spectrum. In Fig.~6, we demonstrate the delocalization of $\chi^{\textrm{in}}$ near the topological phase transition, where the localization length $\xi$ jumps abruptly from a finite value to a value comparable to the system size $L$ at the critical point $\alpha^2=\tilde{\mu}^2+\Delta_0^2$. In the figure we use $R=25$~nm to make $\tilde{\mu}$ appreciable, so that $\alpha\neq\Delta_0$ at the transition. While numerics suggest a transition at $\alpha\approx 0.5218$~meV, the above formula evaluates to $\alpha \approx 0.5220$~meV showing again an excellent agreement.

\begin{figure}
\centering
\includegraphics[width=0.45\textwidth]{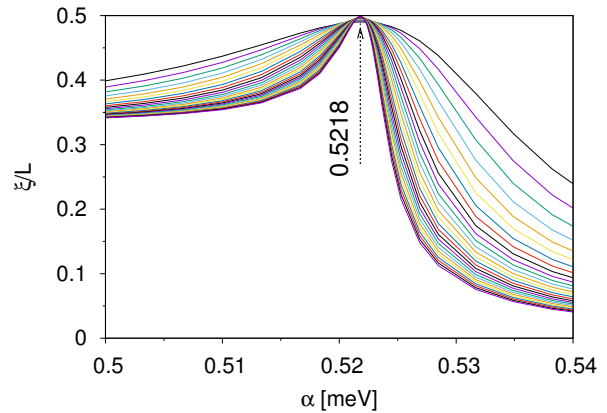}
\caption{The localization length (defined as the inverse participation ratio over half of the space, $\xi^{-1} = \int_0^{L/2} {\rm d}r\, [r \Phi^\dagger(r) \Phi (r)]^2$) of the lowest eigenstate in $l=0$ sector as a function of the exchange energy, for $m=m_e$, $\mu=0$, $\Delta_0=0.5$~meV, $\varphi=0$, $r_0=0$, $R=25$~nm, and $n=2$. The curves in different colors correspond to different system sizes, from $L=1\mu$m to $L=7\mu$m in step of $200$~nm. The dotted arrow is a guide to the eye, denoting the convergence point fitted as $\alpha \approx 0.5218$~meV.}
\end{figure}

\begin{figure}
\centering
\includegraphics[width=0.45\textwidth]{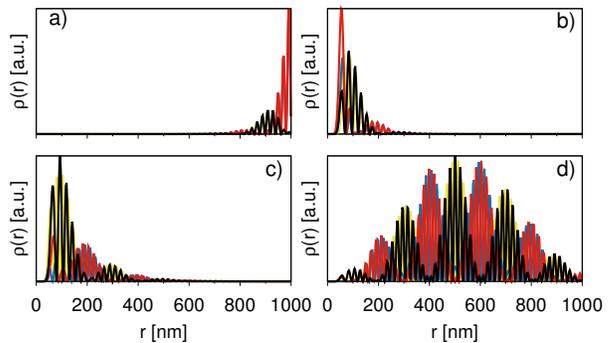}
\caption{Probability densities of the lowest four eigenstates in $l=7$ sector in Fig.~2(c). (a) A subgap state localized at the outer boundary, $r=pR$. (b-c) Subgap states localized near $r=0$. (d) An extended state.}
\end{figure}

\section{Properties of subgap states}

In this appendix, we study the subgap states in the quasiparticle spectra. As typical examples, in Fig.~7 we plot the probability densities of the lowest four eigenstates in $l=7$ sector in Fig.~2(c). The state with the lowest energy belongs to the almost-flat band of the states localized at the outer boundary of the skyrmion, including the Majorana $\chi^{\textrm{out}}$ in the $l=0$ sector. The next two low-lying states belong to the set of localized states near $r=0$. Compared with the MBS $\chi^{\textrm{in}}$, these states are shifted in space towards the outer of the system, which may be pictured semiclassically as due to a ``centrifugal force'' associated with their finite angular momenta. This is a general feature of the subgap states. The highest-energy state lies in the quasi-continuum, with its wave function extending throughout the system. The structures of the subgap states in Fig.~2(e) are qualitatively the same as those in Fig.~2(c).

\begin{figure}
\centering
\includegraphics[width=0.45\textwidth]{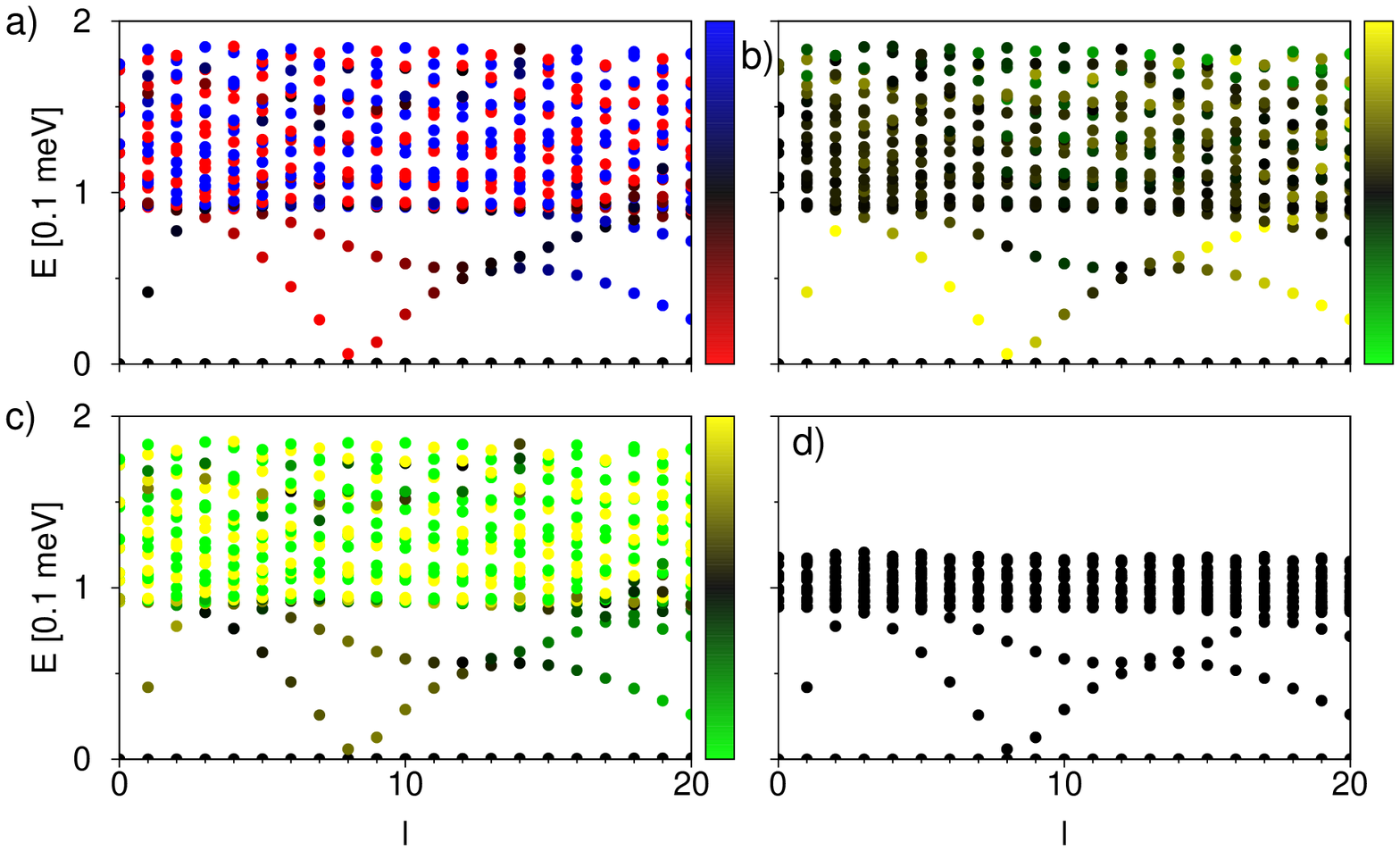}
\includegraphics[width=0.45\textwidth]{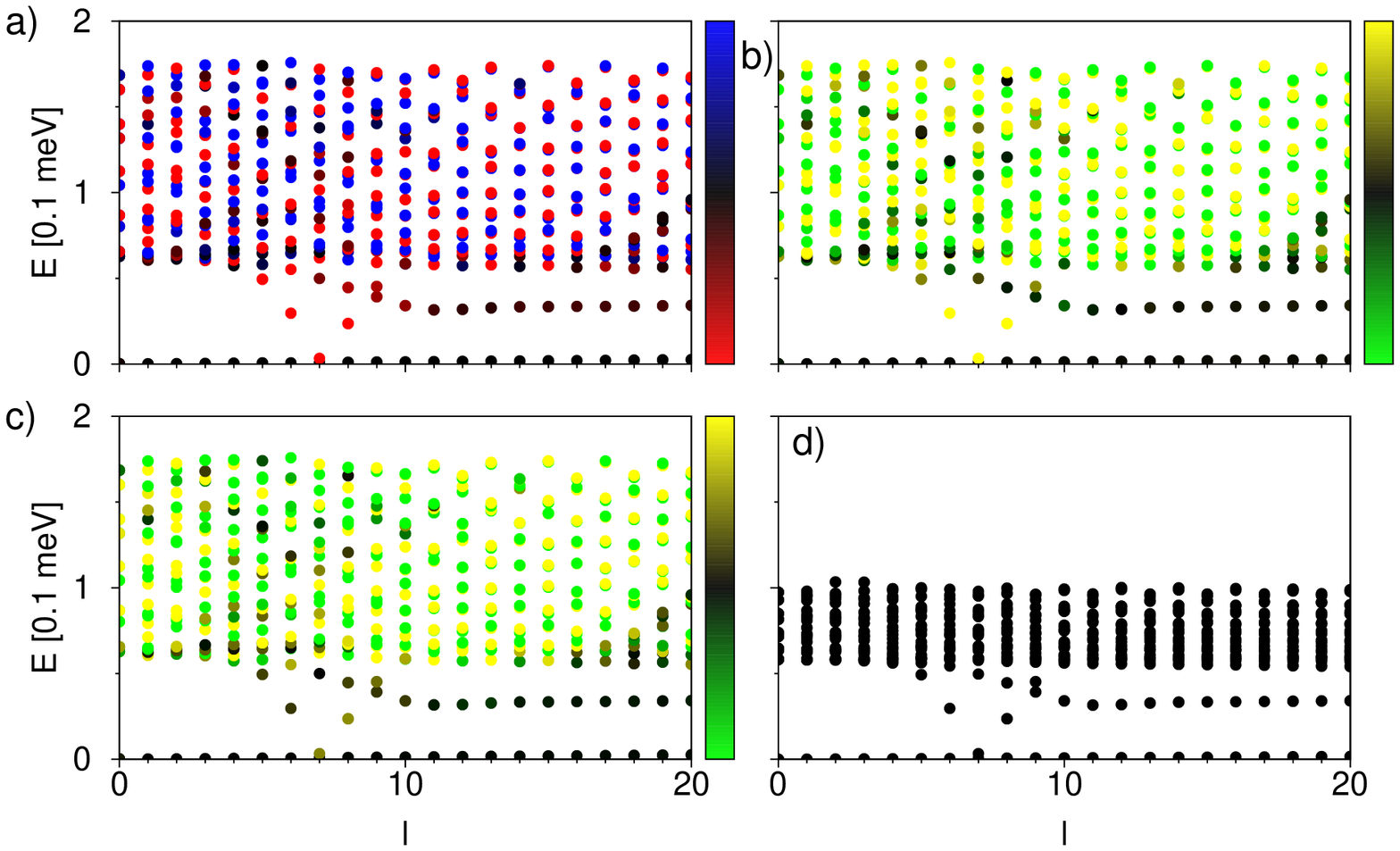}
\caption{Charge and spin expectation values, shown in color as compared with the color bars aside, of the subgap states in Fig.~2(c) (upper 4-panel block) and Fig.~2(e) (lower 4-panel block). The color bar represents a range of $\langle -0.1,0.1\rangle$ for both charge and spin, with full polarization corresponding to $\pm 1$. (a) Mean charge. (b) Mean spin along $\hat{z}$. (c) Mean spin projection along the local skyrmion spin $\hat{N}$. (d) Spectrum for the doubled system size $L=2\,\mu$m, with other parameters unchanged. In all these figures, the upper limit on the energy is a numerical artifact, due to a limited number of eigenstates returned by the diagonalization routine. The true spectrum extends to much higher energies.}
\end{figure}

To gain a better understanding of the subgap states, we have analyzed their charge and spin expectation values, along with those of the extended states in the quasi-continuum. The results are shown in Figs.~8(a)(b)(c). We find that the subgap states near the core of the skyrmion tend to be polarized in both charge and spin, while those near the outer boundary are almost unpolarized. On the other hand, the extended states can be both polarized and unpolarized. In considering TQC utilizing the MBS near the core of the skyrmion, a method discriminating the MBS from the low-lying subgap states is in need. We propose that this can be achieved experimentally by applying an electric field or a magnetic field that couples to the finite charge or spin polarization of the subgap states. 

We also consider the situation where there is a constant phase $\theta_0$ in the skyrmion spin texture,
\begin{equation}
\label{addphase}
\hat{N}(\mathbf{r})=\big(\sin f\cos (n\theta+\theta_0), \sin f\sin (n\theta+\theta_0), \cos f\big).
\end{equation}
In the presence of the extrinsic SOI $H_{so}$, $\theta_0$ is the relative angle between the SOI vector and the local skyrmion spin. In this case, we find that the locations and energies of the subgap states are sensitive to $\theta_0$ values (not shown), but those of the MBS are independent of $\theta_0$.

Lastly, we show in Fig.~8(d) the effect of system size on the spectrum. The figure shows that the subgap states are, unlike the extended states, insensitive to the change of the system size, which indicates that they are states localized by the skyrmionic texture itself  and are inherent to the system under study.


\begin{thebibliography}{99}
\bibitem{footnote2}
The off-diagonal $f''(r)$ term is a consequence of Hermiticity of the Hamiltonian and in our model of a piece-wise linear function $f(r)$ affects only the boundary conditions, which we find by integrating Eq.~(\ref{y7}) over infinitesimal widths near the boundaries. Singularities in $f''(r)$ result in kinks in the Majorana wave function, for example, at $r=r_0$ we have $\tilde{u}_{\uparrow,\downarrow}(r_0^+)=\tilde{u}_{\uparrow,\downarrow}(r_0^-)$ and $\tilde{u}'_{\uparrow,\downarrow}(r_0^+)-\tilde{u}'_{\uparrow,\downarrow}(r_0^-)=\pm\pi\tilde{u}_{\downarrow,\uparrow}(r_0)/2R$, with $r_0^{\pm}=r_0\pm 0^+$.
\end{thebibliography}

\end{document}